\documentclass[prb,preprint]{revtex4}
\usepackage{amssymb}
\usepackage{amsmath}
\usepackage{bbm}
\usepackage[dvips]{color}

\newcommand{\mtin}[1]{\mbox{\tiny {#1}}}
\newcommand{\ca}[1]{{\cal #1}}

\newcommand{\sfrac}[2]{{\textstyle\frac{#1}{#2}}}
\newcommand{\E}{\mathrm{e}}

\newcommand{\myep}{\varepsilon}

\newcommand{\neu}[1]{{#1}}

\begin{document}
\bibliographystyle{apsrev}
\title{Smoothing of Singular Legendre Transforms in Renormalization Group Flows}
\author{C. Husemann}\email{c.husemann@thphys.uni-heidelberg.de}
\author{M. Salmhofer}\email{m.salmhofer@thphys.uni-heidelberg.de}
\affiliation{Institut f\"ur theoretische Physik, Universit\"at Heidelberg,
Philosophenweg 19, 69120 Heidelberg, Germany}


\begin{abstract}
We consider $O(N)$--symmetric potentials with a logarithmic singularity in the second field derivative. This class includes BCS and Gross Neveu potentials. Formally, the exact renormalization group equation for the Legendre transform of these potentials seems to have ill-defined initial conditions. 
We show that the renormalization group equation for the local potential has well-defined initial conditions and that the logarithmic singularity is smoothed rapidly in the flow. Our analysis also provides an efficient method for numerical studies.
\end{abstract}

\maketitle

\section{Introduction}
\neu{In quantum field theory and quantum statistical mechanics, 
bosonic $O(N)$ models originate naturally from microscopic fermionic models as the effective low-energy models for 
order parameter fields, like Cooper pairs 
or spin operators. As such, they play a central role in the 
analysis of symmetry--breaking phenomena.  
Technically, they arise via the} introduction of auxiliary boson fields $\phi$ coupling to composite fermion fields with a Gaussian integral (Hubbard-Stratonovich transformation).
\neu{The correlation function of the composite fermionic 
``order parameter fields'' can then be expressed as functions
of the correlations of the $\phi$, 
and the integration over the fermionic fields yields 
an new action $G_0(\phi)$ for the bosonic fields $\phi$.
There are situations where the resulting action $G_0$ is not 
localized enough, so that the fermionic degrees of freedom
need to be kept even at the lowest scales,  
but there is a large class of models 
where studying $G_0$ is justified at low enough energies.}
  
On the mean--field level, a nonvanishing expectation value of the Hubbard-Stratonovich field 
$\phi$ signals symmetry breaking. In the full theory, fluctuations need to be taken into account (and can strongly change or even invalidate the mean-field result). 
 
The functional renormalization group (RG)\neu{\cite{Wilson1975,WegnerHoughton73,FeldmanTrubowitz91,WetterichTetradisBerges,salmhoferbook}}
is a very useful tool for studying such fluctuation effects: \neu{
it defines a flow of effective actions $G_s$, with initial condition
given by the potential $G_0$, as a function of a scale 
parameter $s \ge 0$. 
In the typical application, $s$ is related to some energy, 
length or temperature scale \cite{Wegner76,TemperatureFlow} that labels
which degrees of freedom are incorporated. 
Here we have taken the convention that the energy scale is a 
decreasing function of $s$ (or the length scale is increasing
in $s$). We remark in passing that the RG method is 
flexible enough to allow for widely varying choices of $s$. 

More and more fluctuation effects are
incorporated as $s$ increases, and the full generating 
function for the correlations is obtained for $s \to \infty$.
The existence of this limit is not obvious. Indeed, 
control over this limit can be considered the solution 
of the model, i.e.\ the construction of a particular model 
of quantum field theory or statistical mechanics. 

There are several different implementations of the RG idea
\cite{Polchinski,TetradisWetterich94,SalmContinuous}, all of which are equivalent on a general level, 
but each with their proper merits and drawbacks
when doing analysis and making approximations. }
The RG differential equation for the generating functional
$\Gamma_s$ for the one-particle irreducible (1PI) 
vertices of $O(N)$ models has had success in a wide range of applications, see Ref.\ \onlinecite{WetterichTetradisBerges} 
for a review. 

It is important to note that the often-studied case of smooth initial potentials, e.g.\ $\phi^4$ potentials,
does not really correspond to a model derived from integrating out the fermionic 
degrees of freedom. In many examples, the second field derivative of the boson potential contains a logarithmic singularity for small fields. The most  prominent example is the BCS theory of superconductivity \cite{BCS}, where the order parameter describes the superconducting gap. The same logarithm in the second field derivative can be seen in the Gross Neveu model \cite{GrossNeveu} and is relevant in the study of mass generation and chiral symmetry breaking in the two--dimensional
situation, where the model is perturbatively ultraviolet renormalizable.

This singularity in the effective potential cannot be regarded as a 
physically irrelevant detail because it implies the persistence of
a symmetry--broken solution down to arbitrarily small values of the interaction strength. 
Indeed, all the familiar formulas of BCS theory would change if the potential were nonsingular. 
(Other features of the fermionic effective potential 
are not well--described by a $\phi^4$ type potential either, 
as discussed below.)

\neu{In this paper, we discuss the role of such initial 
singularities in the RG flow. Let $W$ be the generating functional of the connected correlation functions.}
Following Ref. \onlinecite{WetterichTetradisBerges} 
we set up the RG flow by multiplying the integrand of the functional integral {for $W$} 
with a
regularizing Gaussian exponential with covariance $c_s=R_s^{-1}$ \neu{to obtain a scale dependent generating functional $W_s$}.
\neu{Here $s$ is the RG scale, which runs from zero to infinity, and $R_s$ is a regulator function chosen such that in the 
limit $s \to 0$, $R_s \to \infty$, so that all fluctuations are suppressed at the beginning, and the generating function 
for the amputated correlation functions is equal to the 
initial action $G_0$. In the opposite limit $s\to\infty$,
$R_{s}\to 0$, so that the regulator disappears and 
formally, the full generating function for the correlations
is recovered (as mentioned above, it is nontrivial to show 
that this limit really exists). }
Taking the Legendre transform of the logarithm of the partition function,
subtracting the regulating Gaussian exponent,
and differentiating, we obtain the 1PI flow equation of a
modified Legendre transform \cite{WetterichTetradisBerges}
\begin{align}\label{eq:NonPerturbativeRGE}
\dot{\Gamma}_s[\phi]= \frac 12 \mbox{Tr} \left[ \dot{c}_s
  \frac{\delta^2 \Gamma_s}{\delta\phi^2} \left( \mathbbm{1} +c_s
    \frac{\delta^2 \Gamma_s}{\delta\phi^2} \right)^{-1}\right] \, .
\end{align}
In comparison with Ref. \onlinecite{WetterichTetradisBerges}, Eq.
(\ref{eq:NonPerturbativeRGE}) originates from a normalized partition
function, that is, a term $\frac{1}{2} \mbox{Tr} \dot{R}_sR_s^{-1}$
is subtracted here. This functional equation is exact,  but in most physically
interesting models,  the functional $\phi \mapsto \Gamma_s [\phi]$
has to be approximated for a direct computation. 

There are two common approximations for the functional
$\Gamma_s[\phi]$. First, $\Gamma_s$ can be expanded in powers of the
fields $\phi$ and truncated at some finite even order. If the local
potential of $\Gamma_s$ contains logarithmic terms in $\phi$, 
an expansion around $\phi =0$ is obviously not possible. 
However, as discussed, the logarithm for
small fields in the second field derivative generically 
ensures a nonvanishing mean field solution
$\phi_{\mtin{M.F.}}$. 
\neu{By changing the expansion point to
$\phi_{\mtin{M.F.}}$, one can avoid the logarithm 
in an expansion in $\phi - \phi_{\mtin{M.F.}}$.}
\neu{However, this expansion can then converge
at most for $|\phi| < | \phi_{\mtin{M.F.}}|$, which 
is very small for weak interactions. 
Even worse, }
for the BCS-model we find that the coefficient of
$(\phi-\phi_{\mtin{M.F.}})^4$ has a negative eigenvalue in the
radial mode. Therefore, requiring stability of the functional
integral, a $\phi^4$ truncation is not feasible in this case. This
problem is not cured by including the six-point function or by a
naive separation of small and large fields. It is, of course, merely a problem 
of the expansion in powers of $\phi-\phi_{\mtin{M.F.}}$, 
since the potential is bounded below. -- We note 
in passing that potentials obtained by the Hubbard-Stratonovich transformation and
fermionic integration 
also do not grow like $|\phi|^4$ at large $|\phi|$ but rather 
like $|\phi|^2$ since the logarithm of the fermionic determinant
grows only linearly in $|\phi|$ at large $|\phi|$. 

The other often-used approximation  is a derivative or gradient
expansion \cite{TetradisWetterich94,Morris,MorrisLectures,ReviewBagnulsBervillier}.
While it is not yet clear under which circumstances such expansions are asymptotic
\cite{DerivativeConvergence}, they have been applied 
successfully to a variety of physical
problems in a renormalization context, see Ref. \onlinecite{WetterichTetradisBerges}
and the references therein.
\neu{A naive application of the derivative expansion meets an
ultraviolet problem for the case of fields $\phi$ originating from
a Hubbard-Stratonovitch transformation, because the fermion 
loops determining the action $G_0$ vanish at large 
Matsubara frequencies. Therefore, a time derivative term
is never really there to smoothen the short-time fluctuations, 
i.e.\ the propagator for $\phi$ has no decay at large frequencies. 
Ultraviolet divergences are only prevented by the decay of 
the higher vertices of the initial action in these frequencies, i.e.\ the decay of the vertices generated
by the $\phi$-dependent terms in 
$\frac{\delta^2\Gamma_s}{\delta\phi^2}$.
When the initial action is the result of an integration 
where the high-frequency modes are integrated
over, e.g.\ in a fermionic representation, this ultraviolet problem 
is absent. The fermionic integration 
over high-frequency modes can be done by 
convergent perturbation theory. \cite{PedraSalmhofer}}

\neu{A further problem is that the status of 
(\ref{eq:NonPerturbativeRGE}) becomes unclear 
in the limit $s\to 0$ if the second field derivative
of the initial interaction potential}
contains a singular term,
such as\  $\frac{\delta^2\Gamma_0}{\delta
\phi^2}[\phi_c]\sim \ln \phi_{c}^2$, 
\neu{when evaluated at a constant field $\phi_c$}. 
Certainly, \neu{if one tried to replace} 
$\frac{\delta^2 \Gamma_s}{\delta\phi^2} $
by $\frac{\delta^2 \Gamma_0}{\delta\phi^2} $ in the inverse in Eq.\  
(\ref{eq:NonPerturbativeRGE}), one would end up with a singularity at
some small, $\phi_c$--dependent $s$.

In the present note we show that this problem is not really there, 
due to the smoothing properties of the RG flow, which become evident
when regarding the flow of the connected, amputated functions instead of the 
1PI vertex functions. We show that the generating function for the 
connected functions is smooth at any $s > 0$ and use this to give 
estimates on the Legendre transform that imply smoothness
of $\Gamma_s$ in $\phi$ for any positive $s$. 
We apply this in two ways. 
First, we can overcome the problem of the seemingly ill-defined 
initial condition simply by the semigroup property of the RG:
performing the fluctuation integral with covariance $c_\myep$
\neu{as a Gaussian convolution for $W_\myep$ and respectively $G_\myep$, 
and then Legendre-transforming,} 
gives a new, smooth, initial condition $\Gamma_\myep$ 
for the generating function of the 1PI vertices. It turns out that
$\frac{\delta^2 \Gamma_\myep}{\delta\phi^2} \sim \log c_\myep$, 
so that $c_\myep \frac{\delta^2 \Gamma_\myep}{\delta\phi^2} $
vanishes as $\myep \to 0$, and hence there is no singularity in the
inverse in Eq.\  (\ref{eq:NonPerturbativeRGE}). 
Second, we use these estimates to show that the differential equation 
for the 1PI vertices holds for any $s>0$, and we give the asymptotic 
behaviour of the solution for small $s> 0$. As one would expect, the 
deviation from the initial condition $\Gamma_0$ is nonuniform in 
$\phi$, which explains the absence of the above-mentioned singularity:
at any $s > 0$,  one can choose $\phi$ so small that  $\Gamma_0[\phi]$ 
is not a good approximation for $\Gamma_s[\phi]$.

Thus the physically important logarithmic singularities 
in the initial condition for the potential do not present
any conceptual problem for the \neu{functional} RG, 
and our method also
provides a practical method to treat such initial conditions, 
\neu{also in the 1PI scheme}.
For simplicity of presentation, we concentrate here on reduced
$O(N)$ models, that is, only on the local potential. The field
theoretical methods and the estimates we use allow, however,
\neu{generalize to the full model: the smoothing 
property of the Gaussian convolution also holds for 
infinite--dimensional Gaussian integrals, and the 
strong decay properties imposed by the RG regulator function
at the beginning of the flow justify perturbation theory.}
In particular, the generalization to include the second order 
of a derivative expansion is straightforward.

Let $\phi=(\phi_1,\ldots \phi_N)\in\mathbb{R}^N$ be a constant field, that is, a vector with $N$ components. For $H \in \mathbb{R}^N$ let $(\phi,H)=\sum_{i=1}^N \phi_i H_i$, 
and denote $\phi^2 = (\phi,\phi) $.
We consider a reduced $O(N)$ model with the 
generating function for the connected correlations
\begin{align}\label{eq:functionalW}
W_s(G_0, H)=\ln \int \frac{\mathrm{d}^N \phi}{(2\pi s)^{N/2}} \exp \Big[ -\frac{\phi^2}{2s} -G_0(\phi)+ (\phi,H)\Big] \, .
\end{align}
The external field $H$ couples linearly to $\phi$. 
The scale dependence $c_s=s\mathbbm{1}_N$ with $s\in [0,\infty)$ is already included in the definition. This particular choice of scale dependence is not essential for the calculations; 
it is chosen for convenience only. 
\neu{The potential $G_0$ is 
$O(N)$--symmetric, so that it can be written as 
$G_0(\phi)=V_0(\rho)$ with $\rho=\frac 12 \phi^2$.
We assume that $V_0$ is smooth away from $\rho =0$
and that  for large $\rho$, $V_0' (\rho) \ge \mbox{const.} >0$.
For small $\rho$, we assume 
\begin{align}\label{eq:V0asym}
V_0(\rho)=V_0(0)+v_1 \rho\ln \rho +v_2 \rho+ \ca{R}(\rho) .
\end{align} 
Here $v_1>0$ and the remainder term ${\cal R}$ satisfies 
${\cal R} (0) = {\cal R}'(0) = 0$, and there is a constant
$K_0 > 0$ such that 
$|{\cal R}'' (\rho)| \le K_0 \rho^{-\alpha}$
with $\alpha < 1$. 
With these assumptions, the function $\exp (- G_0(\phi)
+ (H,\phi))$ is integrable uniformly in $H$, hence the limit 
$s \to \infty$ of (\ref{eq:functionalW}) exists by the 
dominated convergence theorem.} 

An important example satisfying these hypotheses is 
the mean-field potential of the BCS model. 
This is the case $N=2$ and 
\begin{align}\label{eq:BCSpot}
V_0(\rho)= \frac{\rho}{g} - \int \mathrm{d} E \; \nu(E) \sqrt{E^2 +\rho} \, ,
\end{align}
if the density of states $\nu(E)$ is regular at the Fermi level $E=0$.  \neu{Here $-g$ is the coupling constant in front of the 
Cooper pair interaction term. The logarithm in  (\ref{eq:V0asym})
is really there, i.e.\ $v_1 > 0$, if $\nu (0) \ne 0$. }

For notational simplicity we have used a unit volume here. In general, the exponent is given 
by $\Omega V_0$, where $\Omega$ denotes the volume, which is taken to infinity in the thermodynamic limit. In this limit, Eq.\ (\ref{eq:BCSpot}) becomes exact for the reduced 
BCS model \cite{Muehlschlegel}. 
In presence of $\Omega$, the factor $|\phi|^{N-1}$ in
the integration measure, 
$\mathrm{d}^N \phi \sim |\phi|^{N-1} \mathrm{d} |\phi| \; \mathrm{d^{N-1}} \omega$\neu{, where $\mathrm{d}^{N-1}\omega$ is the integration measure of the $(N-1)$ dimensional sphere,} 
is not relevant for the following discussion, 
\neu{because  all other parts of the exponent get multiplied by 
$\Omega$. }

The effective potential $\Gamma_s(\phi)=\gamma_s(\phi)-\frac{\phi^2}{2s}$, where $\gamma_s$ is the Legendre transform of $W_s$, is again $O(N)$-symmetric and we write $\Gamma_s(\phi)=U_s(\rho)$
\neu{(recall that $\rho = \frac12 \phi^2$)}. 
Denoting differentiation with respect to the scale $s$ by a dot and differentiation with respect to $\rho$ by a prime we obtain the RG equation 
\begin{align}\label{eq:UFlowEq}
\dot{U}_s=\frac{1}{2} \left[ \frac{(N-1)U_s'}{1+s U_s'} +
  \frac{U_s'+2\rho U_s''}{1+s\big[U_s'+2\rho U_s''\big]}\right]
\end{align}
for the effective (local) potential \cite{WetterichTetradisBerges}, which can also be derived by inserting constant fields in Eq. (\ref{eq:NonPerturbativeRGE}). In this sense $U_s$ is the lowest order of a derivative expansion. Formally, the initial condition is posed in the limit $s\to 0$, where $U_0(\rho)=V_0(\rho)$, which seems to lead to the \neu{vanishing-denominator-}problem discussed before
\neu{because $U_0'(\rho) = v_1 \ln \rho + v_1 + v_2 + O (\rho^{1-\alpha}) \to -\infty$ as $\rho \to 0$. }
\neu{Of course, $W_s$ is convex by Jensen's inequality, 
and hence the Legendre transform cannot diverge
at any finite $\rho$. 
In the following we show the more specific statement that, due to the smoothing effects of the RG transformation, the denominators are strictly positive, 
and we give sharp bounds for their behaviour as $s \to 0$. }

\section{The RG as a smoothing operator} 
\neu{We introduce the effective action}
\begin{align}\label{eq:EffectiveAction}
G_s(\xi) = - \ln \int\frac{\mathrm{d}^N \phi}{(2\pi s)^{N/2}} \exp \Big[ -\frac{\phi^2}{2s} -G_0(\phi+\xi)\Big]
\end{align}
such that $W_s(G_0,H)= \frac{\xi^2}{2s} -G_s(\xi)$ with $\xi=s H$. 
By $O(N)$ symmetry we can write $G_s(\xi)=V_s(\zeta)$ with $\zeta=\xi^2/2$. 
The structure of (\ref{eq:EffectiveAction}) is
\begin{align}
G_s(\xi) 
=
- \ln \left(\mu_s * \E^{-G_0}\right)(\xi)
\end{align}
where $*$ denotes convolution  and $\mu_s$ is the Gaussian measure with covariance $s$
\neu{(the integral exists by the above-mentioned properties of 
$G_0$)}. 
\neu{For $s \to 0$, $\mu_s$ tends to a Dirac measure, so 
the convolution gives $\E^{- G_0}$ in that limit. }
The convolution with a Gaussian measure is a standard example of a smoothing operator
\cite{Hoermander}, 
so this already implies that in spite of the singularities in derivatives of $G_0$, 
$\mu_s * \E^{-G_0}$ is smooth, even analytic in $\phi$ for any $s > 0$. 
This can be seen explicitly from $(\mu_s * f) (\xi) = \int f(x) \mathrm{d} \mu_s (x - \xi)$, 
and understood in a physical analogy by noting that the RG flow defined in 
(\ref{eq:EffectiveAction}) is a heat flow with time parameter $s$, 
whose solution is smooth for any positive time $s > 0$. 

Therefore we can avoid the singular initial condition altogether
by using the semigroup property \cite{salmhoferbook} of Gaussian integration: 
let $\myep>0$, then for all $s > \myep$
\begin{align}
G_s(\xi) 
=
- \ln \left(\mu_{s-\myep} * \E^{-G_\myep}\right)(\xi) .
\end{align}
Or in terms of the unamputated connected functions with a shifted scale
\begin{align}
W_{s}(G_\myep,H) = W_{s+\myep} (G_0, \frac{s}{s+\myep}H) +\frac{H^2}{2}\frac{s\myep}{s+\myep} \, ,
\end{align}
for all $s>0$, i.e.\
\begin{align}\label{eq:NonPertRedW2}
W_{s}(G_\myep,H) = \ln \int \frac{\mathrm{d}^N\phi}{(2\pi s)^{N/2}}\;
\exp \Big[ -\frac{\phi^2}{2s} -G_{\myep}(\phi) + (\phi,H) \Big] \, .
\end{align}
We find $\lim_{s\to \infty} (W_s(G_{\myep},H) -W_s(G_0,H) )=\myep\frac{H^2}{2}$, that is, the functions $W_s(G_0,H)$ and $W_s(G_{\myep},H)$ coincide in the limit $s\to\infty$ up to an explicit term. 
The RG flow of the (modified) Legendre transform remains unchanged but the advantage is now that the initial condition of Eq. (\ref{eq:UFlowEq}) is given by $G_{\myep}(\phi)=V_{\myep}(\rho)$, which is smooth. In the remainder of this section we compute $V_{\myep}$ 
\neu{and give bounds on its derivatives}. 
$V_{\myep}'$ has no logarithmic divergence in  $\phi$ for arbitrarily small $\myep>0$, 
and it provides a well-defined starting point for integrating (\ref{eq:UFlowEq}).

\neu{To begin, we collect some properties of $V_0$ 
that follow from (\ref{eq:V0asym}) and the assumptions
on the remainder term ${\cal R}$ stated there, namely that, 
loosely speaking, the behaviour of $V_0$ is 
that of $v_1 \rho \ln \rho $ for small $\rho$. 
By our assumptions  and integration in $\rho$,
\begin{equation}
|{\cal R}'' (\rho)| \le \frac{K_0}{\rho^\alpha}, 
\quad
|{\cal R}' (\rho)| \le \frac{K_0}{1-\alpha}\rho^{1-\alpha}, 
\quad
|{\cal R} (\rho)| \le \frac{K_0}{1-\alpha}\rho^{2-\alpha}
\end{equation}
with $\alpha<1$. It follows immediately that 
\begin{equation}
|V_0' (\rho) - v_1 \ln \rho |
\le v_1 + |v_2|  + \frac{K_0}{1-\alpha}\rho^{1-\alpha},
\end{equation}
which is much smaller than $|v_1 \ln \rho|$ for small 
enough $\rho$, and 
\begin{equation}
|V_0'' (\rho) - \frac{v_1}{\rho} |
\le \frac{K_0}{\rho^\alpha}
\end{equation}
which is again much smaller than $\frac{v_1}{\rho}$ 
for small enough $\rho$ because $\alpha < 1$. 
The properties of $\rho \mapsto v_1 \ln \rho$ and 
an easy approximation argument then imply that
there is an interval $(0,2\rho_0]$ 
on which the derivative $V_0'$ of the initial potential 
is negative, the map $\rho \to |V_0' (\rho)|$
is decreasing and the maps $\rho \to  \rho  |V_0' (\rho)|^k$, $k=1,2$
are increasing. Moreover, on this interval $|V''_0 (\rho)| \le \frac{c''}{\rho}$
where $c''$ is a constant. In particular we can choose $\myep$ so small that 
$\myep  |V_0' (\myep)| < 0.1$. 
For reasons of brevity, we do not give the detailed values
of the constants as functions of $v_1,v_2,K_0$ and $\alpha$ here. 
}

We split the analysis of  $G_\myep$ in two cases \neu{distinguished by the value of $\zeta=\frac{\xi^2}{2}$}. 

\noindent
{\em Case 1: $\zeta\le \myep\ll 1$. }
We change integration variables to $\hat{\phi}=(\phi+\xi)/\sqrt{\myep}$, subtract $V_0(0)$ in the exponential, and expand the exponential of $V_0(0)-V_0(\hat{\rho} \myep)$\neu{, where $\hat{\rho}= \frac{\hat{\phi}^2}{2}$}. Then perturbation theory for small $\myep$ yields
\begin{align}\nonumber
e^{-G_{\myep}(\xi)}&= e^{-V_0(0)}\Big[ 1- \myep( v_1 \ln \myep +v_2)\frac 12 (N+J^2) \\ &+\myep v_1 T(J) +\ca{O}((\myep\ln \myep)^2) \Big]\, ,
\end{align}
where $J=H\sqrt{\myep}=\frac{\xi}{\sqrt{\myep}}\in [0,\sqrt{2}]$ and
\begin{align}
T(J) =  e^{- \frac{J^2}{2}} \int\frac{\mathrm{d}^N \phi}{(2\pi)^{N/2}}\; \rho \ln \rho e^{-\frac{\phi^2}{2} + \phi J}  = \tilde{T}(\sfrac{J^2}{2})\, .
\end{align}
The function $T(J)$ and all its derivatives with respect to $J$ are bounded on the interval $J\in[0,\sqrt{2}]$. Likewise the higher order terms and their derivatives with respect to $J$ can be estimated. That is, although $\myep$ can be arbitrarily small, $V_{\myep}$ contains no logarithms of the field anymore. Additionally we obtain for the derivatives  
\begin{align}\nonumber
V'_{\myep}(\zeta) &= v_1 \ln \myep +v_2 +  v_1 \tilde{T}'(\sfrac{\zeta}{\myep}) + \ca{O}(\myep(\ln \myep)^2)\\ \label{eq:estimatePerturbation}
V''_{\myep}(\zeta) &= \frac{v_1}{\myep} \tilde{T}''(\sfrac{\zeta}{\myep}) + \ca{O}((\ln \myep)^2) \, .
\end{align}

\medskip\noindent
{\em Case 2:  $\myep< \zeta \le \rho_0$. }
We perform the integral  (\ref{eq:EffectiveAction}) 
by the saddle point method 
\neu{(because we are analyzing $G_s$ for $s=\myep$, 
$s$ is substituted by $\myep$ in (\ref{eq:EffectiveAction}))}. 
The stationarity condition for the negative exponent 
$
S(\phi) =  
\frac{(\phi - \xi)^2}{2 \myep} + V_0 \left(\frac{\phi^2}{2}\right)
$
in the integrand of Eq. (\ref{eq:EffectiveAction}) is 
\begin{align}\label{eq:stationary}
\frac{\partial S}{\partial \phi_i }
=
\frac{1}{\myep}
\left[
\phi_i \left(1 + \myep V_0'\left(\sfrac{\phi^2}{2}\right)\right)
- \xi_i
\right]
=0
\end{align}
for all $i$.
We first assume that there is a stationary point $\phi^*$ and denote 
$\rho^* = (\phi^*)^2/2$. Then (\ref{eq:stationary}) implies
\begin{align}
(\phi^* - \xi)^2 
=
2 \myep^2 \rho^* V_0'(\rho^*)^2
\end{align}
and
\begin{align}\label{eq:rhostar}
\rho^* \left(1 + \myep V_0'(\rho^*)\right)^2 
= \zeta .
\end{align}
The left hand side of (\ref{eq:rhostar}) is monotonically increasing 
in $\rho^*\in [\myep,2\rho_0]$ by our hypotheses on \neu{the potential} $V_0$. Thus a unique solution
$\rho^* \in [\zeta,2\zeta]$ of (\ref{eq:rhostar}) exists. 
There is no solution in the interval $[0,\myep]$ since $\zeta> \myep$. For larger fields there is no solution since $V_0'(\rho)$ becomes positive eventually, so that Eq. (\ref{eq:rhostar}) would imply $\rho^*<\zeta\le\rho_0$, and because $\myep$ is small. 
Given $\rho^*$, the unique solution of (\ref{eq:stationary})
is, by $O(N)$ invariance of $V_0$,  
$\phi^* = \sqrt{2 \rho^*} \, \frac{\xi}{|\xi|}$.
Thus $S$ has a single stationary point. By (\ref{eq:rhostar}),
and because $V_0'(\rho^*) < 0$,
\begin{align}
0 \le \rho^* - \zeta
\le
\rho^*
\left(
2 \myep |V_0' (\rho^*) |
+
\myep^2  V_0' (\rho^*) ^2
\right)\, .
\end{align}
Because $|V_0'|$ is decreasing and $\rho^* \ge \zeta \ge \myep$, 
this implies
\begin{align}
0 \le \rho^* - \zeta
\le
\rho^*
\eta \left(2 + \eta \right)
\le 3 \rho^* \eta
\end{align}
with $\eta = \myep |V_0' (\myep) |$, hence
\begin{align}
\zeta \le \rho^* \le
\frac{\zeta}{1-3 \eta} \, .
\end{align}
We thus have the estimate
\begin{align}
|V_0'(\rho^*) - V_0' (\zeta) |
&\le 
(\rho^* - \zeta )\; \sup\limits_{r \in [\zeta,\rho^*]} |V_0''(r)|
\nonumber\\
&\le 
(\rho^* - \zeta )\; \frac{c''}{\zeta}
\le 
(\rho^* - \zeta )\; \frac{c''}{\rho^* (1-3 \eta)} 
\nonumber\\
&\le 
c'' \frac{3 \eta}{1 - 3 \eta} \, .
\end{align}
These bounds imply that all  eigenvalues of the Hessian 
\begin{align}
H_{ij}
=
\frac{\partial^2 S}{\partial \phi_i \partial \phi_j}
=
\frac{1}{\myep} \delta_{ij} (1 + \myep V_0'(\rho))
+
\phi_i\phi_j V_0''(\rho)
\end{align}
are positive and of order $\myep^{-1}$ at $\phi^*$. 
Thus $\phi^*$ is \neu{the unique} minimum of $S$ and a standard saddle point
analysis\cite{BenderOrszag}  applies:
\neu{all contributions from $\phi$ not in a neighbourhood
of the minimum are suppressed exponentially for small
$\myep$,
as are the corrections to the Gaussian integral around
the saddle point. }
The Gaussian integral around the saddle point gives $(2\pi)^{N/2} D^{-1/2}$, 
where $D = \det H$.
It gives only subleading contributions since the factor $\myep^{N/2}$ 
in $D^{-1/2}$ is canceled by the normalization factor 
$s^{-N/2} = \myep^{-N/2}$ of Eq. (\ref{eq:EffectiveAction}). 
Therefore in case 2,
\begin{align}\nonumber
V_{\myep}(\zeta)&=V_0(\zeta)+ \ca{O}(\myep)\\ \nonumber 
V_{\myep}'(\zeta)&= V_0'(\zeta) + \ca{O}(\myep\ln \myep)\\ \label{eq:estimateSaddle}
\zeta V_{\myep}''(\zeta)&= \ca{O}(1)  \, .
\end{align}
Combining Eqs. (\ref{eq:estimatePerturbation}) and (\ref{eq:estimateSaddle}) from the two cases,
we find that in $V_{\myep}(\zeta)$ the logarithm of the field $\zeta$ is replaced by the logarithm of $\max\{\myep,\zeta\}$. Therefore, the RG flow starting at $s = \myep$, 
and with with initial condition $\Gamma_\myep = V_{\myep}$,
is well defined. 

As we have just shown, perturbation theory for small $\zeta<\myep$ allows us to calculate $V_{\myep}$ to arbitrary precision. 
This result can easily be extended to  non-reduced models because in general, the regularization 
\neu{$c_{\myep}$} provides an infrared regularization, which justifes perturbation theory for small enough \neu{$\myep$}.
For reasons of brevity, we have only outlined the saddle point argument that estimates the 
difference of $V_\myep$ and $V_0$ for $\zeta>\myep$. This argument can easily be made
into a proof, and it also extends to the non-reduced situation, again by noting that the 
infrared regularization together with the smallness of $\myep$ provide rigorous control over
the saddle point expansion.

\section{The RG differential equation at small $s$ and $\phi$}
Shifting the initial condition of the flow as described in the last section is an exact procedure and approximations become necessary only for the calculation of the new initial condition (at least for non--reduced models). But the question remains whether one can find a less indirect way of showing
that the RG equation (\ref{eq:UFlowEq}) for the local potential $U_s$ is well defined at all $s > 0$ if the initial potential contains logarithmic terms. In this section we study the asymptotic solution of the RG equation for small RG scales $s$ and small field squares $\rho = \frac12 \phi^2$. 
As explained below, 
the argument is not solely based on (\ref{eq:UFlowEq}), but requires the bounds derived in the last
section as an a priori input.

In a first step, we assume that the denominators and also $\rho U_s''$ in Eq. (\ref{eq:UFlowEq}) do not contribute to the leading asymptotic solution. Then the flow equation becomes a partial wave equation $\dot{U}_s (\rho)= \frac{N}{2} U_s'(\rho)$, which is solved by the backward propagating wave \begin{align}\label{eq:asymptoticSol}
U_s(\rho)=U_0(\rho+\sfrac{N}{2}s) \, .
\end{align}
If we knew that Eq. (\ref{eq:asymptoticSol}) also provides the asymptotic behaviour for the 
derivatives with respect to $s$ and $\rho$,  we could easily justify the assumptions we just made:
the denominators for small $\rho$ and $s$ contribute only to order $\ca{O}(s\ln s)$, and $\rho U_s''$ is bounded by a constant for small $\rho$. However, asymptotic expressions cannot simply be
differentiated, hence regularity of the derivatives of the local potential cannot  be assured by this argument. The natural procedure starting from the RG equation would now be to differentiate Eq. (\ref{eq:UFlowEq}) with respect to $\rho$. This allows to determine the asymptotic solution and to verify the above assumption for $U_s'$, provided that a regularity assumption is made on $U_s''$.
Another differentiation allows to do the same for $U_s''$, given a suitable hypothesis on $U_s'''$, and so on. 
To avoid an infinite proliferation, it suffices to have a priori bounds for 
$U_s'(\rho)$ and $\rho U_s''(\rho)$ for small $s$ and $\rho$. 
We have already derived such bounds directly from the functional integral in the previous section, 
\neu{and use them now to prove the asymptotic correctness
of  (\ref{eq:asymptoticSol}).}

For the Legendre transformation of $W_s$ we denote the inverse of the maps $\frac{\partial W_s}{\partial H_i}(H) \mapsto \phi_i$ by $\tilde{H}_i(\phi)=\frac{\partial \Gamma_s}{\partial \phi_i} + \frac{\phi_i}{s}$. Using $W_s(H)= \frac{\zeta}{s}-V_s(\zeta)$ we find the connection between the derivatives of the effective action $V_s(\zeta)$ and the local potential $U_s(\rho)$
\begin{align}\label{eq:dULT}
U'_s(\rho) = \frac{V_s'(\tilde{\zeta}(\rho))}{1-sV_s'(\tilde{\zeta}(\rho))} \, ,
\end{align}
where $\tilde{\zeta}(\rho)= s^2\frac{\tilde{H}(\phi)^2}{2}$ is determined by $
\tilde{\zeta}(\rho)= \rho/{\big[1-sV_s'(\tilde{\zeta}(\rho))\big]^2} $.
Combining the estimates of $V'_s(\zeta)$ obtained in Eqs. (\ref{eq:estimatePerturbation}) and (\ref{eq:estimateSaddle}) we arrive at the estimate $|V_s'(\zeta)| \le c \ln (\max \{s,\zeta\})$
for small $s$ and $\zeta$ and a constant $c\in \mathbb{R}$. Using Eq. (\ref{eq:dULT}) this gives $|U_s'(\rho)|\le c \ln s$ with another constant $c$. Similarly, $|V_s''(\zeta)|\le c (\max\{\zeta, s\})^{-1}$ implies $|U_s''(\rho)|\le c \rho^{-1}$ asymptotically for small $\rho$ and $s$. Therefore, Eq. (\ref{eq:asymptoticSol}) is the asymptotic solution of the RG equation.

\section{Conclusion}
We have shown that it is possible to apply the functional RG to
initial conditions given by  potentials with a logarithmic singularity in their second field derivative, 
\neu{because the RG flow smoothes out these logarithms sufficiently fast.}  One might think that a rapid change of the effective local potential near the singularity might cause numerical difficulites, but our arguments also provide a method to
calculate the flow at small $s$ efficiently and with arbitrary precision. 
                                            
We have restricted our analysis to reduced models to bring out the main points in a simple way, 
but it can be generalized to include the second order of a derivative expansion. For example, the $Z_0$ and $Y_0$ functions (see Ref. \onlinecite{WetterichTetradisBerges} for standard notation) diverge with $\rho^{-1}$ and $\rho^{-2}$ respectively for the BCS model. As shown here for the local potential, regularized functions $Z_{\myep}$ and $Y_{\myep}$ can be obtained by a derivative expansion of the effective action at scale $\myep$. 
\neu{Moreover, as explained above, 
the smoothing argument  is completely general, 
that is, it can be used to prove a similar statement 
to the full theory.}

As already remarked in the beginning, potentials with singularities are not academic 
examples, but arise in important physical situations and have important effects. 
The results described here will therefore be useful in going beyond $\phi^4$-type
approximations of these potentials, to obtain a more quantitative theory.

A natural question is whether our analysis also applies to more singular 
initial conditions. It is straightforward to extend our proofs to potentials $V_0$ 
whose derivative diverges as a power of $\log \rho$ for $\rho \to 0$. 
This case includes, in particular, a $(\log \rho)^2$ singularity, which 
occurs in the study of superconductivity of two-dimensional Fermi systems 
with Van Hove singularities.

We acknowledge financial support from DFG research unit FOR 723.

\bibliography{singularLT}
\end{document}